# Inverse Projection Representation and Category Contribution Rate for Robust Tumor Recognition

Xiao-Hui Yang, Li Tian, Yun-Mei Chen, Li-Jun Yang, Shuang Xu, and Wen-Ming Wu

**Abstract**—Sparse representation based classification (SRC) methods have achieved remarkable results. SRC, however, still suffer from requiring enough training samples, insufficient use of test samples and instability of representation. In this paper, a stable inverse projection representation based classification (IPRC) is presented to tackle these problems by effectively using test samples. An IPR is firstly proposed and its feasibility and stability are analyzed. A classification criterion named category contribution rate is constructed to match the IPR and complete classification. Moreover, a statistical measure is introduced to quantify the stability of representation-based classification methods. Based on the IPRC technique, a robust tumor recognition framework is presented by interpreting microarray gene expression data, where a two-stage hybrid gene selection method is introduced to select informative genes. Finally, the functional analysis of candidate's pathogenicity-related genes is given. Extensive experiments on six public tumor microarray gene expression datasets demonstrate the proposed technique is competitive with state-of-the-art methods.

——————————— ◆ ———————————

## 1 INTRODUCTION

With the rapid development of gene chip technology, we can quickly and accurately acquire tumor gene expression microarray data, which have strong ability to measure expression levels of thousands of genes simultaneously. Analyzing and interpreting these gene data can provide aid for tumor early diagnosis on the level of molecular biology [1]. Therefore, effective analysis of microarray gene expression data techniques has attracted much attention in recent years. Microarray gene expression data, however, have the characteristics of small samples (patients), high dimensions (thousands of genes) and high redundancy [2], which impose a challenge to tumor classification.

Microarray gene expression data-based tumor classification mainly consists of clustering [3] and classification [4]. For the characteristic of small sample size, classifier design is still an active and challenging issue for tumor classification [5-8]. Khan et al. [5] developed a method of classifying cancers to specific diagnostic categories based on their gene expression signatures using artificial neural networks. Furey et al. [6] used support vector machine (SVM) to analysis both classification of the tissue samples and give exploration of the data for mis-labeled or questionable tissue results. Shi et al. [7] proposed an improved diagonal discriminant analysis with sparse constraint for tumor classification. Liu et al. [9] proposed a tumor classification based on robust principal component analysis (PCA) and SVM. However, these methods are mostly based on statistical learning theory and need training process to determine model parameters. Recently, deep-learning based classification methods have been proved effective for recognition. However, its success usually relies on big data, complex net structure and advanced hardware.

Sparse representation is a sparse coding technique based on an over-completed dictionary without learning. Sparse representation based classification (SRC) was originally proposed by Wright et al. for face recognition [10]. Xu et al. [11] proposed an integrated sparse representation-based face recognition method, which artificially enlarged training set by constructing symmetry virtual face samples. Our previous work [12] proposed an inverse projection based pseudo-full-space representation classification (PFSRC) for face recognition by focusing on exploiting complementary information among existing available samples rather constructing auxiliary training samples. Recently, SRC has attracted much attention from bioinformatics [13-17]. Hang et al. [13] applied SRC in tumor classification by interpreting gene expression data. Zheng et al. [14] made use of singular value decomposition to learn a dictionary and then classified gene expression data of tumor subtypes based on SRC. Gan et al. [15] improved and generalized [14] by adding a weighted matrix. Khormuji et al. [16] proposed a SRC based tumor classification method, which used geometrical structure of data. Gan et al. [18] used latent low-rank representation (Lat_LLR) for extracting salient features from the original tumor data before SRC. The success of SRC depends on enough training data of the same category. For tumor

————————————————

- X.-H. Yang is with Data Analysis Technology Lab, Institute of Applied Mathematics, School of Mathematics and Statistics, Henan University, Kaifeng 475004, China. E-mail: xhyanghenu@163.com
- L.Tian is with School of Mathematics and Statistics, Henan University, Kaifeng 475004, China.. E-mail: zihaibeitl@163.com
- Y.-M. Chen is with Department of Mathematics, University of Florida, Florida USA. E-mail: yun@ufl.edu
- L.-J, Yang is with School of Mathematics and Statistics, Henan University, Kaifeng 475004, China. E-mail: yanglijun@henu.edu.cn
- S. Xu is with School of Mathematics and Statistics, Xi'an Jiaotong University, Kaifeng 475004, China. E-mail: henuxs@foxmail.com
- W.-M. Wu is with School of Mathematics and Statistics, Henan University, Kaifeng 475004, China. E-mail: wenmingWu55@163.com

classification, however, it is difficult to acquire so many labeled samples. Zhang et al. [19] indicates that the discrimination ability of SRC will be reduced when there is a small disturbance on representation error. It is meaningful to improve the effect of tumor classification if one can tackle these problems of SRC.

On the other hand, there are many irrelevant, redundant and noisy genes and small set of informative genes. It is believed that more reliable cancer classification results will be achieved based on the informative genes. Ranking methods [1, 4, 20] are promising and attractive because their simplicity and stability. Least absolute shrinkage and selection operator (LASSO) [21, 22] is an embedded method, which uses predictor performance as the objective function to evaluate the selected informative gene subset. Hybrid gene selection methods [23, 24] reach more reliable performance by effectively combining complementary strengths from different methods [25]. There are other methods, of course, can be used for gene selection. Wu et al. [26] applied sparse linear discriminant analysis to gene selection. Dai et al. [27] presented an attribute selection method based on fuzzy gain ratio under the framework of fuzzy rough set theory. Cadenas et al. [28] applied fuzzy random forest and feature selection fuzzy random forest with embedded capacity to tumor classification. RPCA technology proposed by Candes was also used for gene selection and achieves good results [9].

Motivated by these works, we propose an inverse projection representation classification (IPRC) to improve the performance of SRC based tumor classification. We restrict our attention to limited training samples and representation without learning. Here, limited training samples mean that there are a small number of training samples (with label) and others are test samples (without label). It is noted that the proposed inverse projection representation focuses on utilizing existing available samples to form the representation space, rather than constructing auxiliary samples by other ways. The main differences between the proposed IPRC and the related works [10-12] are as follows. (1) IPRC focuses on a completely opposite projection way to [10] and presents a novel classification criterion to match the inverse representation and fulfill classification, which similar to [12]. (2) [11] also mentioned inverse representation, while the projection way is different because different applications. [11] represented each training face sample of a category with a test sample, training samples of the other categories and their symmetry virtual face samples, while IPRC focuses on the available test sample space. The classification criterion of [11] is the same with [10]. (3) Our previous work [12] proposed an inverse projection for face recognition. However, [12] projects each training sample into pseudo-full space because face images have important complementary between samples, while it not suitable for gene expression data. More importantly, a statistical measure is constructed to quantify the stability of representation-based classification methods.

The remainder of this paper is organized as follows. The presented robust tumor classification based on two-stage gene selection and IPRC is stated in Section 2. Extensive experimental results are shown in Section 3. Finally, Section 4 concludes the paper.

## 2 METHODOLOGY

### 2.1 Inverse Projection Representation Based Classification

#### 2.1.1 Sparse Representation Based Classification

Suppose $X = [x_1, \cdots, x_{|X|}]$ are training samples, Let $|X|$ be labeled training samples in total of $c$ categories. SRC [10] assumes a test sample $y_r$ can be represented as,

$$y_r = \sum_{i=1}^{|X|} x_i \alpha_{r,i}, \tag{1}$$

where $i = 1, 2, \cdots, |X|$, $r = 1, 2, \cdots, k$, $\alpha_{r,i} \in R$ is the coding coefficients. Let $\alpha_r = [\alpha_{r,1}, \cdots, \alpha_{r,i}, \cdots, \alpha_{r,|X|}]^T$, $l_1$-norm with the following Lagrangian formulation is often adopted.

$$\hat{\alpha}_r = \arg\min_{\alpha_r} \left\{ \|y_r - X\alpha_r\|_2^2 + \lambda \|\alpha_r\|_1 \right\}. \tag{2}$$

The classification criterion of SRC is as follows,

$$e_r^j = \|y_r - X_j \delta_j(\hat{\alpha}_r)\|_2^2, j = 1, \cdots, c, \tag{3}$$

where $\delta_j : R^n \to R^n$ is a characteristic function that selects coefficients associated with the $j$-th category. For $x \in R^n$, $\delta_j(\hat{\alpha})$ is a vector whose only nonzero entries in $\alpha$ that are associated with category $j$. A test sample is classified into the category with the minimal reconstruction error.

Next, the stability of SRC will be analyzed. Suppose $X_1$, $X_2 \in R^{m \times n}$, which come from two different categories. For a sample $y_r$ from one category, a coefficient vector and error can be calculated: $\alpha_i = \arg\min_\alpha \|y_r - X_i \alpha\|_2$ and error $e_i = y_r - X_i \alpha_i$, $i = 1, 2$.

Suppose the difference between $X_1$ and $X_2$ is a small disturbance $\Delta(X_1) = X_2 - X_1$, which results in $y_r$ has a small disturbance $\Delta(y_r)$. The error can be calculated

$$\varepsilon = \max\left\{ \frac{\|\Delta(X_1)\|_2}{\|X_1\|_2}, \frac{\|\Delta(y_r)\|_2}{\|y_r\|_2} \right\} \le \frac{\varphi_{|X|}(X_1)}{\varphi_1(X_1)}, \tag{4}$$

where $\varphi_1(X_1)$ and $\varphi_{|X|}(X_1)$ are the largest and the smallest singular values of $X_1$, respectively. Refer to [19], the relationship between $e_{r,1}$ and $e_{r,2}$ can be written as,

$$\frac{\|e_{r,2} - e_{r,1}\|_2}{\|y_r\|_2} \le \varepsilon(1 + \kappa_2(X_1))\min\{1, m-n\} + O(\varepsilon^2), \tag{5}$$

where $\kappa_2(X_1) = \|X_1\|_2 \cdot \|(X_1^T X_1)^{-1} X_1^T\|_2$ is the $l_2$-norm

conditional number of $X_1$. It is obviously that the bigger the similarity of $X_1$ and $X_2$ is, the smaller the difference between $e_{r,1}$ and $e_{r,2}$ is. Eq. (5) demonstrates that misclassification is easy to happen and the classification is unstable when $e_{r,1}$ is similar to $e_{r,2}$.

### 2.1.2 Inverse Projection Representation

Suppose there are few training samples (with label) per category, and the others are test samples (without label). In this case, SRC doesn't work well. Therefore, an IPR is proposed to obtain a more stable representation by exploring test sample space.

The projection way of IPR is opposite to that of sparse representation. Let $Y = [y_1, \cdots, y_k]$ is test sample space, where $k$ expresses the number of test samples. Each training sample $x_i$ can be represented by all test samples.

$$x_i = \gamma_{i,1} y_1 + \cdots + \gamma_{i,r} y_r + \ldots + \gamma_{i,k} y_k, \quad (6)$$

where $\gamma_{i,r} \in R$ are representation coefficients. Let $\gamma_i = [\gamma_{i,1}, \cdots, \gamma_{i,k}]^T$ represents coefficient vector, $x_i$ can be rewritten as $x_i = Y\gamma_i$. And then all training samples $X = [x_1, \cdots, x_{|X|}]$ can be linearly represented as follows.

$$X = Y\gamma, \quad (7)$$

where $\gamma = [\gamma_1, \cdots, \gamma_{|X|}]$ is the coefficient matrix.

Zhang et al. [19] indicates that it is the collaborative representation, but not sparsity, that plays the essential role for classification in SRC. Moreover, it is also proved that the $l_1$-norm can be replaced by $l_2$-norm, which can achieve similar classification results but with significantly lower complexity. Therefore, $l_2$-regularized constraint is used in IPR model.

$$\hat{\gamma}_i = \arg\min_{\gamma_i} \left\{ \|x_i - Y\gamma_i\|_2^2 + \lambda \|\gamma_i\|_2^2 \right\}, \quad (8)$$

where $\lambda$ is a regularization parameter.

The analytic solution of matrix form $\hat{\gamma}$ with regularized least square about Eq. (7) is easily and analytically derived as

$$\hat{\gamma} = (Y^T Y + \lambda I)^{-1} Y^T X. \quad (9)$$

As a result, IPR can be more easily implemented than standard sparse representation. What we emphasize is that, the representation space may be enlarged by using test samples, especially there are a small number of training samples per category.

It is easy to notice that the latter focuses on the column coefficients before test samples, rather than row coefficients of training samples for the former. The different projection way makes the IPR is less sensitive to the number of training samples than that of sparse representation.

The feasibility of the proposed IPR can be further analyzed as follows. Similar to [19], for the simplicity of analysis, the regular term in Eq. (8) is removed and then the representation becomes a least square problem,

$$\hat{\gamma}_i = \arg\min_{\gamma_i} \|x_i - Y\gamma_i\|_2^2.$$

Let $x_i^j$ represent a training sample $x_i$ belongs to category $j$, which can be represented by the test sample space based on IPR. Similar to standard sparse representation and without cause confusion, suppose $Y^j$ denotes test sample subspace belong to the same category with $x_i$, the associated representation $\hat{x}_i^j = \sum_j Y^j \delta_j(\hat{\gamma}_i)$ is actually the perpendicular projection of $x_i$ onto the test sample full space $Y$. The reconstruction error by each category $e_j = \|x_i^j - Y^j \delta_j(\hat{\gamma}_i)\|_2^2$ is used for classification. It can be readily derived by

$$e_j = \|x_i^j - Y^j \delta_j(\hat{\gamma}_i)\|_2^2 = \|x_i^j - \hat{x}_i^j\|_2^2 + \|\hat{x}_i^j - Y^j \delta_j(\hat{\gamma}_i)\|_2^2.$$

Obviously, it is the amount $e_j^* = \|\hat{x}_i^j - Y^j \delta_j(\hat{\gamma}_i)\|_2^2$ that works because $\|x_i^j - \hat{x}_i^j\|_2^2$ is a constant for all categories.

Denoted by $\chi_j = Y^j \delta_j(\hat{\gamma}_i)$ and $\hat{\chi}_j = \sum_{m \neq j} Y^m \delta_m(\hat{\gamma}_i)$, $m = 1, \cdots, c$, $m \neq j$, since $\hat{\chi}_j$ is parallel to $\hat{x}_i^j - Y^j \delta_j(\hat{\gamma}_i)$, we can readily have

$$\frac{\|\hat{x}_i^j\|_2}{\sin(\chi_j, \hat{\chi}_j)} = \frac{\|\hat{x}_i^j - Y^j \delta_j(\hat{\gamma}_i)\|_2}{\sin(\chi_j, \hat{x}_i^j)},$$

where $(\chi_j, \hat{\chi}_j)$ is the angle between $\chi_j$ and $\hat{\chi}_j$, and $(\chi_j, \hat{x}_i^j)$ is the angle between $\chi_j$ and $\hat{x}_i^j$. Finally, the representation error can be represented by

$$e_j^* = \|\hat{x}_i^j - Y^j \delta_j(\hat{\gamma}_i)\|_2^2 = \frac{\sin^2(\chi_j, \hat{x}_i^j) \|\hat{x}_i^j\|_2^2}{\sin^2(\chi_j, \hat{\chi}_j)}. \quad (10)$$

Eq. (10) shows that by using IPR, when judging if $x_i^j$ has a strong correlation with a test sample, we need not only consider if $\sin(\chi_j, \hat{x}_i^j)$ is small and also consider if $\sin(\chi_j, \hat{\chi}_j)$ is large. Such a "double checking" makes the representation effective and robust.

### 2.1.3 Category Contribution Rate

It can be observed that the conventional classification criteria, reconstruction error, doesn't work for IPR. Since the representation dictionary is unlabeled test samples. Hence, a classification criterion, CCR, is constructed to match the proposed IPR and complete classification, which is called IPRC.

**Definition 1 (Category Contribution Rate, CCR)** For a test sample $y_r$, the contribution rate $C_{j,r}$ of $y_r$ for the $j$-th category can be calculated by Eq. (11).

$$C_{j,r} = \frac{1}{s_j} \sum_i \left( \frac{\delta_j(\{|\gamma_{i,r}|\})}{\sum_i \{|\gamma_{i,r}|\}} \right), i = 1, \cdots, |X|, \quad (11)$$

where $j = 1, 2, \cdots, c$, $r = 1, 2, \cdots, k$, $s_j$ denotes the number of $j$-th category training samples. For eliminating effects of training sample size may differ in different categories, the projection coefficient vector of every category before $y_r$ is normalized by summing up itself and solving the average. And then the CCR matrix $[C_{j,r}]$, $j = 1, \cdots, c$, $r = 1, \cdots, k$ for all test samples is got. The larger the CCR is, the higher the correlation between each test sample and every category is. A test sample $y_r$ is classified into the category with the maximal contribution rate.

$$m_r = \arg\max_{j \in \{1, \cdots, c\}} (C_{j,r}). \quad (12)$$

By this means, categories of all test samples are obtained simultaneously and classification can be completed.

Comparing Eqs. (3) and (11), one can see that the difference between reconstruction error and CCR lies in that the latter focuses on the coefficients before each test sample rather than the former focuses on those of training samples. Experiments will be shown in Subsection 3.3.2.

### 2.1.4 Stability Analysis of IPRC

**Theorem (Classification Stability of IPRC)** *Suppose $x_i, x_j$ are $i$-th and $j$-th training samples, and the relationship $x_i$ and $x_j$ is $x_j = x_i + \Delta(x_i)$, where $\Delta(x_i)$ is a disturbance of $x_i$. Based on the test samples $Y$, the IPRs of $x_i, x_j$ are as follows: $x_i = Y\gamma_i$, $x_j = Y\gamma_j$, where $\gamma_i$ and $\gamma_j$ are representation coefficients for $x_i$ and $x_j$, respectively. Let $\Delta(Y)$ represents the disturbance corresponding to $\Delta(x_i)$. If*

$$\varepsilon = \max\left\{\frac{\|\Delta(x_i)\|_2}{\|x_i\|_2}, \frac{\|\Delta(Y)\|_2}{\|Y\|_2}\right\} \leq \frac{\varphi_k(Y)}{\varphi_1(Y)},$$

*and $\sin(\theta) = \rho_{LS}/\|x_i\|_2 \neq 1$, where $\rho_{LS} = \|Y\gamma_{LS_i} - x_i\|_2$, $\gamma_{LS_i} = \arg\min_{\gamma_i} \|x_i - Y\gamma_i\|_2$, then*

$$\frac{\|\gamma_j - \gamma_i\|_2}{\|\gamma_i\|_2} \leq \varepsilon \left\{\frac{2\kappa_2(Y)}{\cos(\theta)} + \tan(\theta)\kappa_2(Y)^2\right\} + O(\varepsilon^2). \quad (13)$$

*where $\kappa_2(Y)$ ($\kappa_2(Y) = \|Y\|_2 \cdot \|(Y^TY)^{-1}Y^T\|_2$, $\kappa_2(Y)^2 = \|Y\|_2^2 \cdot \|(Y^TY)^{-1}\|_2$) is the $l_2$-norm conditional number of $Y$, and $\theta$ is angle between $x_i$ and its projection vector on $Y$.*

**Proof.** In order to discuss the value of $\frac{\|\gamma_j - \gamma_i\|_2}{\|\gamma_i\|_2}$, we need to find the relationship between $\gamma_i$ and $\gamma_j$. Let $\gamma_i(t)$ is continuously differentiable for all $t \in [0, \varepsilon]$, where $\gamma_i = \gamma_i(0)$ and $\gamma_j = \gamma_i(\varepsilon)$. Let $\gamma_i(t)$ do the Taylor expansion at $t = 0$: $\gamma_i(t) = \gamma_i(0) + t\gamma_i'(0) + O(t^2)$. We have $\gamma_j = \gamma_i + \varepsilon\gamma_i'(0) + O(\varepsilon^2)$ when $t = \varepsilon$. Then

$$\frac{\|\gamma_j - \gamma_i\|_2}{\|\gamma_i\|_2} = \varepsilon \frac{\|\gamma_i'(0)\|_2}{\|\gamma_i\|_2} + O(\varepsilon^2). \quad (14)$$

In order to obtain $\|\gamma_i'(0)\|_2$, similar to Theorem 5.3.1 in [29], one can construct $(Y + tf)^T(Y + tf)\gamma_i(t)$, where $f = \Delta(Y)/\varepsilon$, then

$$(Y + tf)^T(Y + tf)\gamma_i(t) = (Y + tf)^T(x_i + t\Delta(Y)\gamma_i(t)/\varepsilon).$$

Let $E = \Delta(x_i)/\varepsilon$, then

$$(Y + tf)^T(Y + tf)\gamma_i(t) = (Y + tf)^T(x_i + tE). \quad (15)$$

In order to bound $\|\gamma_i'(0)\|_2$, one can take the derivative of Eq. (15) and set $t = 0$, $f^T Y\gamma_i + Y^T f\gamma_i + YY^T\gamma_i'(0) = Y^T E + f^T x_i$ i.e.,

$$\gamma_i'(0) = (Y^T Y)^{-1} Y^T (E - f\gamma_i) + (Y^T Y)^{-1} f^T(x_i - Y\gamma_i). \quad (16)$$

By singular value decomposition theorem [29], we have $rank(Y + tf) = k$ for all $t \in [0, \varepsilon]$, where $\|\Delta(Y)\|_2 \leq \varphi_k(Y)$ ($\varphi_k(Y)$ is the largest singular value of $Y$). Then

$$\|f\|_2 = \|\Delta(Y)/\varepsilon\|_2 \leq \varphi_k(Y) \leq \|Y\|_2,$$

and $\|E\|_2 = \|\Delta(x_i)/\varepsilon\|_2 \leq \|x_i\|_2$.

By substituting Eq. (16) result into Eq. (14), taking norms, the inequality can be obtained,

$$\frac{\|\gamma_j - \gamma_i\|_2}{\|\gamma_i\|_2} \leq \varepsilon \left\{ \|Y\|_2 \cdot \|(Y^T Y)^{-1} Y^T\|_2 \cdot \left(\frac{\|x_i\|_2}{\|Y\|_2 \|\gamma_i\|_2} + 1\right) + \frac{\rho_{LS}}{\|Y\|_2 \|\gamma_i\|_2} \cdot \|Y\|_2^2 \cdot \|(Y^T Y)^{-1}\|_2 \right\} + O(\varepsilon^2).$$

Since $Y^T(Y\gamma_i - x_i) = 0$, $Y\gamma_i$ is orthogonal to $Y\gamma_i - x_i$, it is also known that $\|x_i - Y\gamma_i\|_2^2 + \|Y\gamma_i\|_2^2 = \|x_i\|_2^2$, then $\|Y\|_2^2 \cdot \|\gamma_i\|_2^2 \geq \|x_i\|_2^2 - \rho_{LS}^2$.

The relationship between $\gamma_i$ and $\gamma_j$ will be

$$\frac{\|\gamma_j - \gamma_i\|_2}{\|\gamma_i\|_2} \leq \varepsilon \left\{ \kappa_2(Y)\left(\frac{1}{\cos(\theta)} + 1\right) + \kappa_2(Y)^2 \frac{\sin(\theta)}{\cos(\theta)} \right\} + O(\varepsilon^2).$$

The conclusion indicates that the distance between $\gamma_i$ and $\gamma_j$ is very small when $x_i$ is similar to $x_j$ (in other words, $Y$ has a small disturbance $\Delta(Y)$). Compared Eq.

(13) with Eq. (5), one can see that coefficients are more sensitive to a small disturbance $\Delta$ than that of reconstruction error. Because, for nonzero residual problems, it is the square of the condition number that measures the sensitivity of coefficients. In contrast, according to Subsection 2.1.1, reconstruction error sensitivity linearly depends on the condition number $\kappa_2(X_1)$. Moreover, it is worth noting that we focus on the column coefficient vector $\gamma_{1,1}, \gamma_{1,2}, \cdots, \gamma_{|X|,1}$ before each test sample when we calculate CCR. However, it has been demonstrated that disturbance will affect row coefficients rather than column coefficients. Moreover, the effect on column coefficients is a positive impact when CCRs of different categories are calculated.

For further quantifying the classification stability of representation-based methods, we propose a statistic measure named as CSI.

**Definition 2 (Classification Stability Index, CSI)** For the representation-based classification methods, suppose $R_{best}^1$ and $R_{best}^2$ are the values of a classification criterion corresponding to the best category and the second best category. The CSI of a test sample is defined to measure the difference between $R_{best}^1$ and $R_{best}^2$. The CSI is normalized as $CSI \in [0,1]$ and is always defined as the ratio of the smaller one and the larger one.

$$CSI = R_{best}^1 / R_{best}^2.$$

For SRC, the CSI is denoted as $CSI_{RE}$, where $R_{best}^1$ and $R_{best}^2$ are the minimal reconstruction error and the second minimal one. While for IPRC, the CSI is denoted as $CSI_{CCR}$, where $R_{best}^1$ and $R_{best}^2$ are the second maximal CCR and the maximal one. The smaller the index is, the better the stability is, the better the representation-based method is. Experiments will be shown in Subsection 3.3.3.

## 2.2 TWO-STAGE HYBRID GENE SELECTION

A two-stage hybrid gene selection method is presented to extract informative genes and to further improve the performance of IPRC for tumor classification.

### 2.2.1 The First Stage-Gene Pre-selection

The first stage, gene pre-selection, aims to primarily select information genes by top-ranked intersection of three filter methods, analysis of variance, ANOVA) [20], signal noise ratio (SNR) [1] and the ratio of between-groups to within-groups sum of squares (BW) [4].

$$G_f = \{G_{ANOVA}\} \cap \{G_{SNR}\} \cap \{G_{BW}\},$$

where $G_f$ is the gene subset based on the first stage, $G_{ANOVA}$, $G_{SNR}$ and $G_{BW}$ are the gene subsets based on analysis of variance, SNR and BW, respectively.

This step primarily picks up candidate genes and reduces the computational complexity. Without loss of generality, the significance level $p = 0.05$ is selected for ANOVA.

### 2.2.2 The Second Stage-Gene Refinement

The second stage, an embedded approach, LASSO-logistic regression [22] is used to perform the gene refinement, which further efficiently refines the smaller subset of candidate information genes.

$$(\hat{\beta}_0, \hat{\beta}) = \arg\max_{\beta_0, \beta} \sum_{i=1}^{N}\left[ y_i\left(x_i \beta^T + \beta_0\right) - \log\left(1 + e^{x_i \beta^T + \beta_0}\right) \right] - \eta \sum_{j=1}^{d} |\beta_j|,$$

where $x_i$ and $y_i$ express the gene expression data and the label of $i$-th sample respectively, and $\beta$ denotes regression coefficient vector and is designed to cope with the case that $y$ follows multinomial distribution.

The penalty parameter $\eta$ in LASSO is selected corresponding to the best classification accuracy on the training set by ten-fold cross-validation. The LASSO model is trained on nine-fold of training set, while the validation is conducted on the other fold training set.

It is worth noting that LASSO is introduced to select gene subset which is used to subsequent classify, that is, the quality of gene subset depends on the classification effect. Therefore, the parameters are selected based on the classification effect rather than those selected by LASSO itself. That is to say, the validation is based on the classification results, which is based on the gene subset selected from the trained LASSO model.

First of all, we do ten-fold cross validation on the training set, and give the average error of ten folds as cross-validation error,

$$Error_i = \frac{1}{10}\sum_{j=1}^{10} err_{i,j},$$

where $Error_i$ is the cross validation error of the parameter $\eta_i$, and $err_{i,j}$, $j=1,\cdots,10$ is the verification error of every fold. It should be noted that the verification error is the classification error on the verification set (that is, the other fold training set).

Then, the parameter $\hat{\eta}$ corresponding to the minimum cross-validation error $Error^*$ is selected as the final parameter of LASSO,

$$\hat{\eta} = \arg\min_{\eta_i \in \{\eta_1, \cdots, \eta_N\}} (Error_i).$$

The gene subset corresponding to the parameter is just the one used for classification.

For the selected parameter $\hat{\eta}$, the corresponding gene subsets $G_j, j=1,\cdots,10$ obtained by ten folds are not exactly the same. Therefore, we compare each fold gene subsets, and finally chose the one corresponding to the minimum error,

$$\hat{G} = \arg\min_{G_j \in \{G_1, \cdots, G_{10}\}} err_{*,j},$$

where $err_{*,j}$ is the verification error for each fold.

### 2.2.3 Two-Stage Hybrid Gene Selection

The reasons of using two-stage hybrid gene selection method which combined filter method with embedded method are as follows. Filter method can provide general solutions for various classifiers because it is independent of any learning algorithm. However, filter methods ignore the interactions between classifiers and may not be suitable for all classifiers. Embedded methods, to some extent, can solve the problems of the filter approach by considering the dependencies on features and classifiers. However, the computational complexity is a major issue, especially can be intractable for large datasets. As the filter method efficiently reduces the size of the gene set, the computational complexity of embedded method becomes acceptable and two methods bring out the best to each other. The framework of the two-stage gene selection method is shown in Fig. 1.

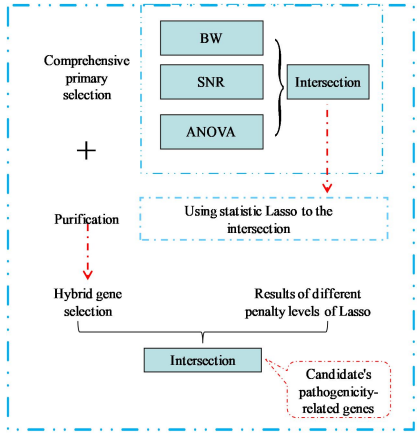

Fig. 1. Flowchart of the two-stage hybrid gene selection.

## 2.3 TUMOR CLASSIFICATION BASED ON TWO-STAGE HYBRID GENE SELECTION AND IPRC

Combined the two-stage hybrid gene section with IPRC, the basic idea of our robust tumor classification algorithm is as follows.

**Input:** Training sample set $X = [x_1, x_2, \cdots, x_{|X|}]$, training label set $L = [l_1, l_2, \cdots, l_{|X|}]$ and test sample set $Y = [y_1, y_2, \cdots, y_k]$.

**Preprocessing:** Standardize the observations (arrays) to have mean 0 and variance 1 across variables. Two-stage hybrid gene selection is applied to $\{(x_i, l_i), i=1,\cdots,|X|\}$ and then is applied to $Y$. And we get samples only with informative genes.

**Classification based on IPRC:**
  Step1. By Eq. (6), the IPR is realized.
  Step2. By Eq. (9), the projection coefficient matrix is got.
  Step3. By normalizing the CCR matrix, relevancies between each test sample and all categories are obtained.

**Output:** By Eq. (12), each test sample can be classified into the category with the maximal CCR.

**Identification of pathogenic genes:** Based on the two-stage hybrid gene selection, the informative genes are selected as the candidate pathogenic subset, whose occurrence number is more than a threshold value.

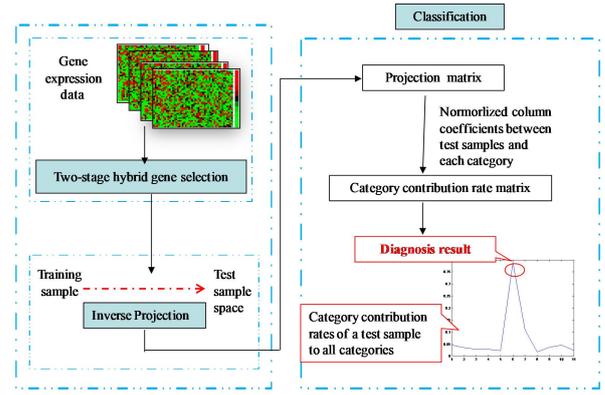

Fig. 2. Flowchart of the two-stage hybrid gene selection and IPRC for robust tumor classification.

Fig. 2 shows the framework of IPRC with two-stage hybrid gene selection for robust tumor classification.

## 3 RESULTS

Experiments are demonstrated on six public gene expression datasets. Six kinds of measures are used to measure the performance of these methods. Accuracy measures the classification performance by using the percentage of correctly classified samples. Sensitivity measures the non-missed diagnosis performance by using the rate of correctly classified positive samples. Specificity measures the non-misdiagnosis performance by using the rate of correctly classified negative samples. For any test, there is usually a trade-off between the sensitivity and specificity. This tradeoff can be represented graphically using a receiver operating characteristic curve (ROC), which is a graphical plot that illustrates the diagnostic ability of a binary classifier system as its discrimination threshold is varied. AUC is just the area under the curve of ROC and is also suitable to binary classification problem [30]. Error reduction rate (ERR) [31] intuitively characterizes the proportion of the errors reduced by switching a method to the other one. Without loss of generality, ten-fold cross-validation ten times is used to test the performance of the algorithms. All experiments are carried out using MATLAB R2016a on a 3.30GHz machine with 4.00GB RAM.

### 3.1 Tumor Data Sets

Six public benchmark cancer microarray gene expression datasets are used to evaluate the performances of our methods: Colon [32], DLBCL [33], SRBCT [5], 9_Tumors [34], 11_Tumors [35] and Leukemia [36]. The first two are binary category datasets and the remaining four are multi-categories datasets. Colon dataset consists of gene expression data of 40 tumor and 22 normal colon tissue samples. The number of genes is 2000. DLBCL dataset consists of gene expression data of diffuse large B cell lymphoma, follicular lymphoma. There are 77 samples, each of which contains 5469 genes. SRBCT dataset consists of small, round blue cell tumors (SRBCT) of childhood. There are 2308 genes in each sample and 83 samples. 9_Tumors dataset consists of gene expression data of nine different human tumor types, such as NSCLC, colon and breast, including 60 samples. Each sample has 5726 genes. 11_Tumors dataset consists of gene expression data of eleven different human tumor types, such as ovary, breast and colorectal. There are 12533 genes in each sample and 174 samples. Leukemia dataset consists of gene expression data of acute myelogenous leukemia, acute lymphoblastic leukemia and mixed-lineage leukemia, including 72 samples. Each sample has 11225 genes.

### 3.2 Parameters selection

#### 3.2.1 Parameter of LASSO Model

Fig.3 gives the cross validation error at different values of parameter $\eta$ on Colon dataset. The abscissa is the number of selected genes, and the ordinate is the cross-validation error. Points on the curve represent parameter values of $\eta$. The dotted arrow and the solid arrow corresponding to the parameters selected by LASSO itself and our proposed method, respectively.

Fig. 3 shows that the final selected parameter $\hat{\eta}=1.20e\text{-}4$ and the number of genes is 143. It is worth noting that the parameter selected by LASSO itself is $6.73e\text{-}2$, and the corresponding gene subset has higher cross-validation error and lower classification accuracy. That is to say, the parameter selected by LASSO itself is not the one we want, because it doesn't get the best classification results. Similarly, Fig.4 gives the results on DLBCL dataset, where the parameter of LASSO is $\hat{\eta}=6.80e\text{-}5$ and the number of selected genes is 270.

Without loss of generality, if multiple folds (greater than one fold) have the same lowest error, one can choose any of them. In this paper, we further calculate the entropy of each fold of gene subset, and select the one corresponding to the lowest entropy. The reason for that is entropy means uncertainty, and the greater the entropy is, the greater the uncertainty is. Fig.6 shows that the validation errors (Fig. 5(a)) and entropies (Fig. 5(b)) of each fold on Colon dataset, where the selected parameter is $\hat{\eta}=1.20e\text{-}4$. The gene subset of the fifth-fold is selected as the final selected gene subset because it has the minimum verification error and entropy.

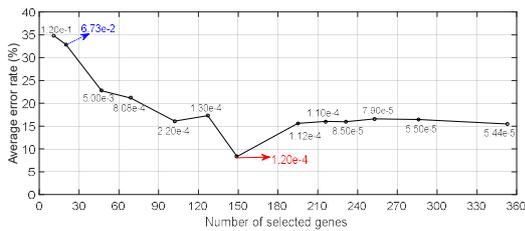

Fig.3. The numbers of genes and average error rate corresponding to different penalty parameters $\eta$ on Colon dataset.

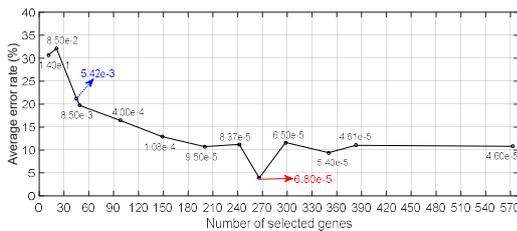

Fig.4. The numbers of genes and average error rate corresponding to different penalty parameters $\eta$ on DLBCL dataset..

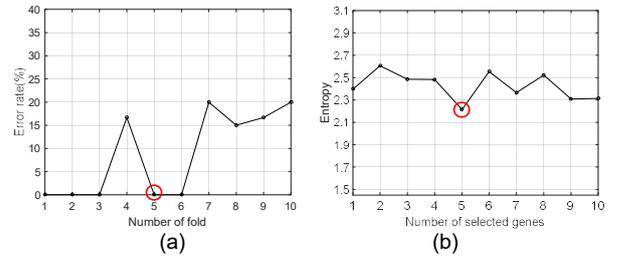

Fig. 5. Classification results of each fold on Colon dataset, (a) error rate, (b) entropy.

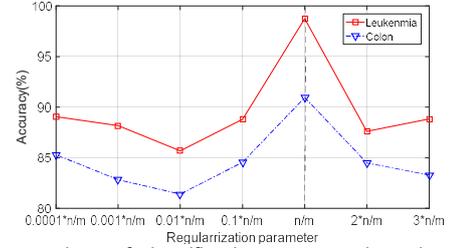

Fig. 6. Comparison of classification accuracy based on different regularization parameter values.

#### 3.2.2 Parameter of Inverse Projection Representation Model

By using the method described in Subsection 2.3, we firstly randomly separate the six datasets into training set and test set. In all experiments, the regularization parameter $\lambda$ is set to $n/m$ in Eq. (8), where $n$ and $m$ are the numbers of test samples and training samples, respectively. On the one hand, the parameter setting is similar to PFSRC [12] and CRC [19]. On the other hand, the regularization parameter $\lambda$ is tested by experiments. Taking binary category dataset (Colon) and multi-category dataset (Leukenmia) as examples, the regularization parameter $\lambda$ are set to $0.0001*n/m$, $0.001*n/m$, $0.01*n/m$, $0.1*n/m$, $2*n/m$, $3*n/m$ and $n/m$, respectively. We select the parameter value corresponding to the optimal classification accuracy. Fig. 6 shows that $\lambda = n/m$ is clearly better than the others.

### 3.3 Results of Tumor Classification Based on IPRC

In this subsection, the performance of the proposed IPRC method is demonstrated. The comparable methods are SVM, SRC and some improved SRC methods. SVM is chosen because SVM [37-38] outperform K-nearest neighbors and neural network in gene expression cancer diagnosis [35]. Although we mainly focus on the proposed projection way and classification criterion, it is believed that improved strategies in SRC methods also can be embedded in our IPPC framework.

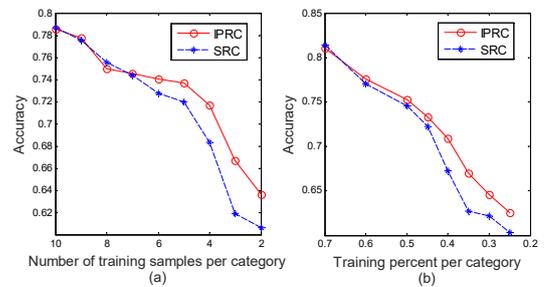

Fig. 7. Comparison of accuracies with decreasing training samples per category. Curves of accuracy versus number of training samples per category on (a) Leukemia dataset and (b) 11_Tumors dataset.

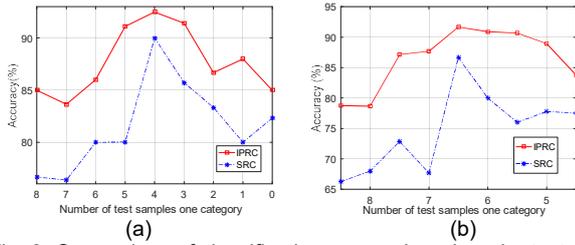

Fig. 8. Comparison of classification accuracies when the test data is not balanced in each category. (a) Colon dataset, (b) Leukenmia dataset.

### 3.3.1 Comparison of Inverse Projection Representation and Sparse Representation

By taking full advantage of the information embedded in test samples, the IPR can relieve the problem of insufficient training samples. The performance of standard representation and the proposed IPR are compared by reducing the number of training samples per category. In order to verify the stability, we perform on two different categories distribution datasets. The Leukemia dataset has a balanced distribution on all categories of sample number, while 11_Tumors has a badly unbalanced sample. For dataset with balanced distribution, the number of training samples per category is reduced from 10 to 2 in Fig. 7 (a). While for dataset with unbalance distribution, the percentage of training sample number per categories is decreased from 70% to 25% in Fig. 7 (b). From Fig. 8, it can be seen that SRC and IPRC reach similar results when the number of training samples is more than 6 per category or percentage is more than 45%. With decreasing the amount of training samples, classification accuracy of SRC will soon lower than IPRC. The results show that IPRC performs more stable than SRC, especially when there are few training samples.

For testing the performance of the proposed IPRC model when the test data is not balanced in each category, the experiments on binary category dataset and multi-category are done. Without loss of generality, binary category dataset (Colon) and multi-category dataset (Leukenmia) are selected as examples. We fix the number of test samples in one category and change that of another category from more to none (zero). Experiments are given in the Fig. 8, which shows that: (1) the category-imbalance does affect the classification results, and the classification accuracies of category-balance is superior to category-imbalance. (2) the optimal classification accuracy is achieved when the numbers of samples are balanced. (3) IPRC has higher accuracies and better stability than SRC either category-balance or category-imbalance.

### 3.3.2 Comparison of Category Contribution Rate and Reconstruction Error

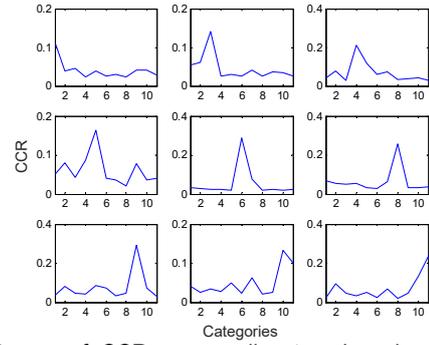

Fig. 9. Curves of CCR versus all categories about nine test samples on 11_Tumors dataset. Test samples will be classified into the category with the only peak.

For verifying the CCR has discrimination power for tumor classification, we randomly select some test samples and calculate the corresponding CCR results across all categories. Some individuals of 11_Tumors are randomly taken as examples, nine test samples (55th, 29th, 116th, 21th, 159th, 66th, 106th, 59th and 131th samples in order) on 11_Tumors dataset. Fig. 9 gives the CCR results of these test samples versus all categories. It can be seen that there is only one peak (the maximum CCR) in every subfigure obviously, which means that we can judge the category of a test sample based on the maximum CCR.

Next, it is demonstrated CCR for IPRC is superior to reconstruction error for SRC. It's worth noting that the more obvious the difference between categories is, the stronger the discrimination ability is, and the better the classification criterion is. Figs. 10 and 11 give the results of the two criterions about some randomly selected test samples in binary category (Colon) and multi-category (11_Tumors) datasets, respectively. The same color

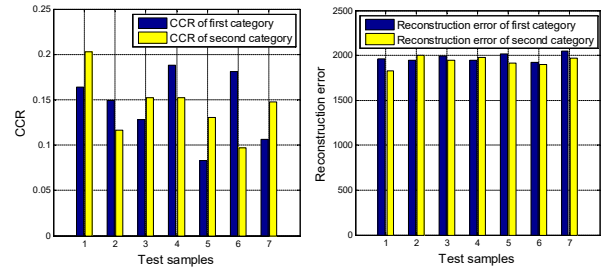

Fig. 10 The comparison of seven samples random selected on Colon dataset. (a) CCR (b) Reconstruction error. The same color histogram expresses the same category.

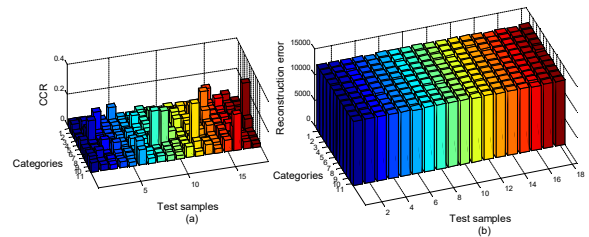

Fig. 11. The comparison on 11_Tumors dataset. (a) CCR (b) Reconstruction error. This 3-dimensional histogram shows the values of two classification criterions of test samples across all classes. The same color expresses the values of the same test sample across all categories.

TABLE 1
CLASSIFICATION RESULTS ON SIX DATASETS

| Methods | Accuracy | Sensitivity | Specificity | AUC |
|---|---|---|---|---|
| Colon dataset | | | | |
| SVM | 85.48 | 72.73 | 92.50 | 0.8080 |
| SRC | 85.48 | 72.73 | 92.70 | 0.8455 |
| IPRC | 88.81 | 87.50 | 90.90 | 0.9250 |
| DLBCL dataset | | | | |
| SVM | 94.09 | 98.28 | 89.47 | 0.8721 |
| SRC | 94.75 | 98.28 | 94.74 | 0.9537 |
| IPRC | 89.82 | 91.37 | 100 | 0.9855 |
| 9_Tumors dataset | | | | |
| SVM | 65.10 | 33.33 | 92.16 | - |
| SRC | 66.67 | 33.33 | 92.16 | - |
| IPRC | 66.67 | 44.44 | 94.11 | - |
| 11_Tumors dataset | | | | |
| SVM | 94.68 | 92.50 | 95.80 | - |
| SRC | 94.83 | 92.59 | 95.92 | - |
| IPRC | 95.00 | 93.10 | 96.27 | - |
| Leukemia dataset | | | | |
| SVM | 96.60 | 94.74 | 94.12 | - |
| SRC | 95.83 | 94.74 | 94.06 | - |
| IPRC | 96.90 | 96.43 | 97.73 | - |
| SRBCT dataset | | | | |
| SVM | 100 | 100 | 100 | - |
| SRC | 100 | 100 | 100 | - |
| IPRC | 100 | 100 | 100 | - |

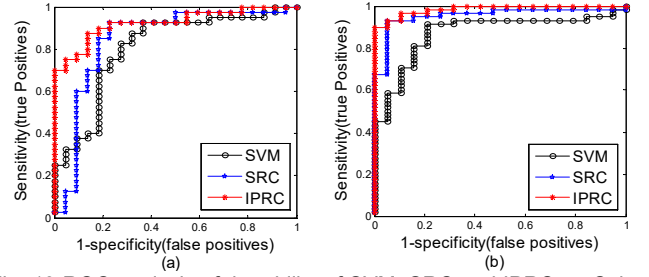

Fig. 13 ROC analysis of the ability of SVM, SRC and IPRC on Colon and DLBCL datasets. Note that on the vertical axis, the scale is from no (0) to complete (1 or 100%) sensitivity. The horizontal axis is a reciprocal scale (1-specificity). The optimum performance of a test is determined either as the highest sum of the specificity and sensitivity to say, the difference between $R_{best}^1$ and $R_{best}^2$ in CCR for IPRC is much bigger than those in reconstruction error for SRC. This further verifies CCR for IPRC has better discrimination ability than reconstruction error for SRC.

### 3.3.4 Results of IPRC-Based Tumor Classification

The performance of IPRC for robust tumor classification is demonstrated in this subsection. For comparison, the results of SVM and SRC are listed under the same experimental environment. For each experiment, we run the ten-fold cross validation ten times and take the means as the final results.

Table 1 and Fig.13 show that IPRC achieves competitive results with highest AUC, which shows IPRC has the best prediction ability among the three classifiers. ROC plot analysis in Fig. 13 has shown that IPRC has the better discrimination ability than SVM and SRC. The accuracy and sensitivity of IPRC are higher than SVM and SRC on Colon dataset. Especially sensitivity of IPRC is 14.77% higher than SVM and SRC, that is, the missed diagnosis rate of IPRC is the lowest. Although the specificity is slightly lower than SVM and SRC. It is worth noting that the patients with acute abdominal pain as main symptoms are susceptible to missed diagnosed in clinical treatment. Hence, high sensitivity and low missed diagnosis rate are indeed needed and helpful for early clinical diagnosis. For DLBCL dataset, the patients will face multiple courses of chemotherapy and great psychological stress if follicular lymphoma is misdiagnosed as diffuse large B-cell lymphoma. Therefore, we want to reduce misdiagnosis as far as possible. IPRC just has the specificity of 100%, which means the rate of misdiagnosing follicular lymphoma as diffuse large B-cell lymphoma is 0%. For multi-categories datasets, Table 1 shows IPRC is superior to SVM and SRC. Moreover, one can also observe that the two sparse representation-based methods, SRC and IPRC, have higher sensitivity and specificity than SVM. As for the acceptable level of accuracy, sensitivity and specificity for a given disease depend on clinical context. For instance, the accuracies of the Colon dataset are 82.73%-90.91%, the AUCs of the same dataset are 84%-93% [9]. Dettling et al. [39] demonstrated similar levels of accuracy of 87.1%. García-Nieto et al. [40] give the sensitivity and specificity of 85.93% and 83.89%. Dang et al. [41] got the sensitivity and specificity of 81.82% and 90.95%. Consequently we expresses the values of a test sample across all categories. According to the overall trend, one can see that, to the same test sample, difference between categories of CCR is much bigger than that of reconstruction error. This shows that the CCR has better discrimination power than construction error. Moreover, for example, the CCR classifies the sixth test sample in Fig. 10 easily, but reconstruction error is hard to discriminate and leads to a wrong classification. The classification stability will be further verified in the following Subsection.

### 3.3.3 Results of Classification Stability

The classification stability is further verified by comparing the quantitative indicator of stability, CSI. Fig.12 shows the CSI of all samples on the six datasets (Colon, DLBCL, SRBCT, 9_Tumors, 11_Tumors and Leukemia in order). The smaller the CSI is, the better the stability is. One can see that $CSI_{RE}$ is almost close to 1 in all subfigures, while the $CSI_{CCR}$ is much smaller. That is can draw the conclusion that the results of IPRC are in the acceptable range.

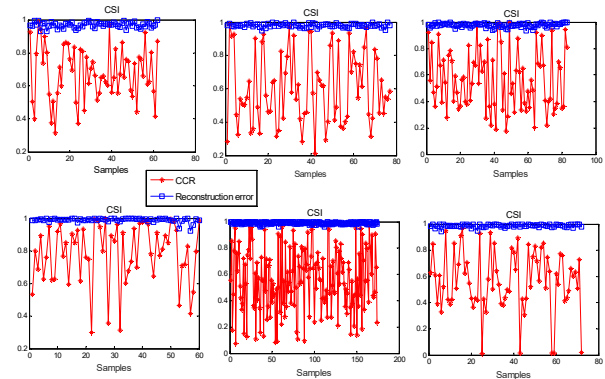

Fig. 12. The curve of CSI versus all samples in ten-fold cross validation. The star line expresses CSIs of CCR in IPRC, and square line expresses CSIs of reconstruction error in SRC.

TABLE 2
COMPARATIVE ERROR RATES ON SRC AND IPRC

| Datasets | Error Rate (%) | | ERR |
|---|---|---|---|
| | SRC | IPRC | |
| Colon | 14.52% | 11.19% | ↓ 22.93% |
| 11_Tumors | 5.17% | 4.02% | ↓ 22.24% |
| Leukemia | 4.17% | 3.10% | ↓ 25.66% |
| 9_Tumors | 33.33% | 33.33% | ↓ 0 |

More intuitively, ERR [31] is introduced by denoting a notion ↓. Table 2 lists the ERR results by switching SRC to IPRC on Colon, 11_Tumors, Leukemia and 9_Tumors datasets. Since the accuracy on SRBCT dataset is 100% for all classifiers, the ERR need not to be calculated. For DLBCL dataset, the ERR doesn't also need to calculate because the classification accuracy is slightly lower than SRC. For instance, since the IPRC reduces the error rate from 14.52% to 11.19%, the ERR is 22.93% [(14.52-11.19)/14.52], suggesting that 22.93% recognition errors can be avoided by using IPRC instead of SRC.

From all these results, one can conclude that IPRC method is feasible and effective for not only binary tumor classification problems but also multi-category tumor classification problems. The reason for IPRC is superior to SVM and SRC may due to the following two facts. Firstly, the number of training sample is small, while SRC and SVM do not consider the information embedded in test data. Secondly, the CCR is more stable to a small disturbance than reconstruction error, which has been validated in Subsections 3.3.2 and 3.3.3.

### 3.3.5 Results of Comparing with Some Improved SRC Methods

The performance of IPRC is also compared with those of some recent SRC-based methods, SRC_Lat_LRR [18], LLE+SR [16], MRSRC [15] and MSRC-SVD [14]. It is worth noting that these compared methods combine SRC with some relatively complex techniques. The classification accuracies of 9_Tumors and 11_Tumors datasets are listed in Table 3, which shows IPRC achieves competitive results and is somewhat slightly higher than MSRC and MRSRC for 9_Tumors dataset. The average classification time over 10 runs of IPRC and SRC are shown in Table 4. Compared with SRC, IPRC needs much less time. Tables 3 and 4 show that IPRC leads to competitive classification results with simple model and low computational complexity.

TABLE 3
ACCURACY OF DIFFERENT METHODS ON 9_TUMORS AND 11_TUMORS

| Experiments | Methods | Dataset | |
|---|---|---|---|
| | | 9_Tumors | 11_Tumors |
| This paper | IPRC | 66.67 | 95 |
| Gan et al.(2014) | SRC_Lat_LRR | 66.67 | 94.83 |
| Khormuji et | LLE+SR | 66.75 | 96.42 |
| Gan et al.(2013) | MRSRC | 60.00 | 95.40 |
| Zheng et al.(2011) | MSRC-SVD | 63.33 | 95.98 |

TABLE 4
AVERAGE CLASSIFICATION TIME ON SIX DATASETS

| | Colon | DLBCL | 9_tumors | 11_tumors | Leukemia | SRBCT |
|---|---|---|---|---|---|---|
| SRC | 0.6556 | 3.8659 | 2.4286 | 25.1199 | 5.2407 | 1.0906 |
| IPRC | 0.0011 | 0.0022 | 0.0045 | 0.0270 | 0.0037 | 0.0019 |

### 3.4 Results of Tumor Classification Based on Two-Stage Gene Selection and IPRC

Classification results of IPRC with and without gene selection are given in this section. Since the accuracy on SRBCT dataset is 100% for all the classifiers, we do the experiments on the other five datasets.

Firstly, we illustrate the necessity of gene selection. Corresponding to Table 1, Table 5 gives classification results based on BW gene pre-selection method and our

TABLE 5
CLASSIFICATION RESULTS ON FIVE DATASETS
(WITH GENE SELECTION)

| Methods | Accuracy | Sensitivity | Specificity | AUC |
|---|---|---|---|---|
| Colon dataset | | | | |
| SVM with BW | 87.10 | 77.27 | 92.50 | 0.7273 |
| SRC with BW | 87.10 | 81.82 | 90.00 | 0.8852 |
| IPRC with BW | 90.48 | 87.50 | 90.90 | 0.8841 |
| IPRC with first stage | 90.95 | 92.50 | 90.90 | 0.9523 |
| DLBCL dataset | | | | |
| SVM with BW | 94.40 | 96.55 | 89.47 | 0.9029 |
| SRC with BW | 94.75 | 98.28 | 94.74 | 0.9610 |
| IPRC with BW | 93.75 | 91.37 | 100 | 0.9819 |
| IPRC with first stage | 94.82 | 92.25 | 100 | 0.9846 |
| 9_tumors dataset | | | | |
| SVM with BW | 66.82 | 33.33 | 94.12 | - |
| SRC with BW | 68.21 | 40.00 | 96.42 | - |
| IPRC with BW | 66.82 | 44.44 | 97.73 | - |
| IPRC with first-stage | 73.55 | 44.44 | 98.09 | - |
| 11_tumors dataset | | | | |
| SVM with BW | 95.00 | 92.50 | 96.29 | - |
| SRC with BW | 94.91 | 92.59 | 95.92 | - |
| IPRC with BW | 95.96 | 96.29 | 99.31 | - |
| IPRC with first stage | 96.18 | 96.29 | 99.31 | - |
| Leukenmia dataset | | | | |
| SVM with BW | 97.22 | 97.37 | 94.12 | - |
| SRC with BW | 96.42 | 97.06 | 94.72 | - |
| IPRC with BW | 98.33 | 97.73 | 100 | - |
| IPRC with first stage | 98.75 | 97.73 | 100 | - |

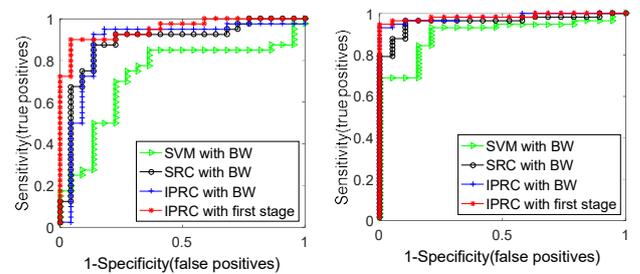

Fig. 14. ROC analysis of the ability of SVM with BW gene pre-selection, SRC with BW gene pre-selection, IPRC with BW gene pre-selection and IPRC with first-stage hybrid gene selection on Colon and DLBCL datasets. The vertical axis is sensitivity. The horizontal axis is 1-specificity.

TABLE 6

CLASSIFICATION ACCURACIES OF OUR METHOD WITH DIFFERENT SELECTION STAGES

| Datasets | Selection methods | | | |
|---|---|---|---|---|
| | Original | LASSO | First stage | Two stage |
| Colon | 88.81 | 90.95(978) | 90.95(389) | 91.90(143) |
| DLBCL | 89.82 | 92.32(2002) | 94.82(828) | 96.07(270) |

two-stage hybrid gene selection method on five datasets. From Table 5, one can see that BW method plays a positive role on all methods. On DLBCL dataset, BW gene pre-selection improves the performance of SVM obviously, but achieves a little improvement on SRC and IPRC. The proposed two-stage hybrid gene selection method can further improve the performance of IPRC, especially for multi-category datasets. Fig. 14 gives the ROC corresponding to Table 5.

Next, the performance of the IPRC based on the proposed two-stage hybrid gene selection will be demonstrated on Colon and DLBCL datasets. Table 6 gives the classification results. There are two reasons for only discussing binary-category classification at LASSO-based gene refinement stage. One is the advantage of IPRC on multi-category datasets has already been proved in Table 5. The other is the fact that Zhang et al. [42] shows that LASSO for multi-category of genetic selection faces great difficulties. Table 6 shows that the classification accuracy increases with decreasing the number of information genes, when we perform the proposed two-stage gene selection.

Next, the performance of gene selection is visualized using principal component analysis. Fig. 15 represents 62 samples consisting of 40 Colon tumor (stars) and 22 normal (squares) using the top three principal components of total 2000 genes, 389 genes based on BW gene selection method and 143 genes based on the proposed two-stage hybrid gene selection method respectively. Fig. 15 (a) shows that a few of the 2000 genes provide classification information and the distribution just looks uniform in each direction. Fig. 15(b) shows that the 389 genes can mostly separate different cancers. Fig. 15(c) also shows the 143 genes have the best separability than those of Figs. 15 (a) and (b). All this suggests that the informative genes based on the two-stage method contain the main classification discriminant information.

Compared with [9], experiments are conducted on Colon and 11_Tumors datasets. In Colon, the IPRC with two-stage hybrid gene selection (91.90%) performs better than the RPCA+LDA+SVM method (90.45%). In 11_Tumors, the IPRC with gene pre-selection performs (96.18%) slightly less than the RPCA+LDA+SVM method (99.34%), that is, IPRC achieves competitive effect although combined with simple gene pre-selection method on multi-category dataset.

### 3.5 Analysis of Candidate's Pathogenic Genes
Apart from obtaining high classification accuracy results, it is also important to identify pathogenicity-related genes, which can be a biomarker of early diagnosis and be helpful to auxiliary diagnosis.

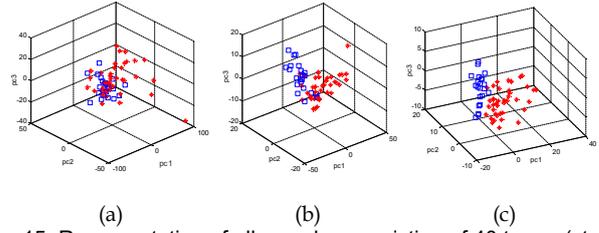

Fig. 15. Representation of all samples consisting of 40 tumor (stars) and 22 normal (squares) on Colon datasets. the top three components of (a) original genes, (b) pre-selected genes and (c) two-stage hybrid selected genes.

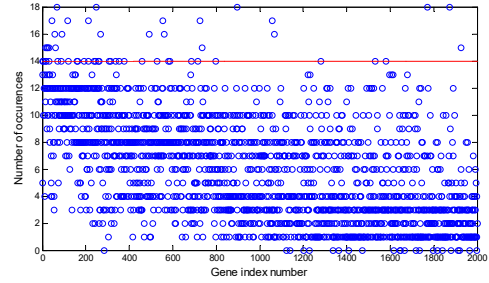

Fig. 16. Number of occurrences versus gene index number on Colon dataset. In general, the more times it occurs, the more important the gene is. The line expresses the threshold of occur frequency.

As shown in Subsection 3.4, candidate's pathogenic genes can be selected by the two-stage hybrid gene selection method based on different penalty levels of logistic regression with LASSO. Firstly, the curve of appearance times versus gene index number is plotted by adjusting the penalty level $\eta$ in LASSO-logistic regression. Fig. 16 illustrates the correlation between every gene and Colon tumor in a degree. Therefore, it can be conjectured that the more it occurs, the more relevant with tumor it is. Then the candidate pathogenic subset contains the genes, which occur more than a threshold value (here, 13 times). At last, the intersection genes over the threshold value are selected as the candidate pathogenic gene subset.

Some genes from the final candidate subset for Colon data are shown in Table 7, which are believed to be closely related to Colon cancer. Gene H08393 has been turned out to be associated with Colon cancer in clinical [43-44]. For further illustration, the related function of these genes is searched in NCBI dataset. For instance, Collagen 11, a heterotrimeric molecule consisting of $\partial 1$, $\partial 2$ and $\partial 3$ chains have role in formation of collagen

TABLE 7

LIST OF THE BEST SUBSET OF SOME GENES FOR COLON DATASET

| Index no.of selectedgenes | Gene accession number | Gene description |
|---|---|---|
| 493 | R87126 | Myosin heavy chainonmuscle (Gallus gallus) |
| 1772 | H08393 | Collagen (XI) chain (Homo sapiens) |
| 249 | M63391 | Human desmin genecomplete cds. |
| 625 | X12671 | Human gene for heterogeneous nuclear ribonucleoprotein (hnRNP) core protein |
| 66 | T71025 | Human (HUMAN);mRNA sequence |
| 1873 | L07648 | Human MXI1 mRNA,complete cds. |
| 897 | H43887 | Complement factor D precursor |

fibrils. *COL*11*A*1, a gene for collagen (H08393), which is normally not expressed in adult colon tissue, has been found to be expressed in colorectal carcinomas. Another collagen gene, COL5A2, normally not expresses but has been found co-expressed with COL11A1 in tumors. HNRNPA1 gene (X12671) encodes a member of a family of ubiquitously expressed heterogeneous nuclear ribonucleoproteins (hnRNPs), which are RNA-binding proteins that associate with pre-mRNAs in the nucleus and influence pre-mRNA processing, as well as other aspects of mRNA metabolism and transport. The protein encoded by this gene is one of the most abundant core proteins of hnRNP complexes and plays a key role in the regulation of alternative splicing. Quantitative alteration of hnRNPA1 may result in facilitation of transformation of colon epithelial cells as a consequence of transcriptional and translational perturbation. Desmin gene (M63391) encodes a muscle-specific category III intermediate filament. Homopolymers of this protein form a stable intracytoplasmic filamentous network connecting myofibrils to each other and to the plasma membrane. Mutations in this gene are associated with desmin-related myopathy, a familial cardiac and skeletal myopathy (CSM), and with distal myopathies.

In order to check the quality of the selection processes, the expression profiles of the final identified genes for the opposite category are analyzed. For comparison, an irrelevant gene chosen randomly is presented. In Fig. 17, the curve with star denotes gene expression levels of 40 tumor samples and the curve with square expresses gene expression levels of 22 normal samples. The line indicates the mean values of gene expression levels in corresponding class. One can see in both cases the mean value of the samples belonging to tumor category differs significantly from the referenced (normal) category. Fig. 18 shows the image of the expression profiles for the two pathogenic genes (H08393 and X12671) and two irrelative genes (M22488 and R72644) in the form of the colormap of jet, where transition from small value to high value corresponds to a shift from low to high expression values of the samples. The vertical axis represents samples (20 tumor samples, 10 normal samples) and the horizontal the genes arranged by index number 1772, 625, 1122, and 1408 respectively. Fig. 18 demonstrate that moderate to high upregulation of H08393 and X12671 and downregulation for other two genes can indicate the presence of Colon. There is a visible difference between samples of the Colon tumor group and the reference one in H08393 and X12671 but similar expression levels in M22488 and R72644, which confirms good performance of the proposed gene selection procedure.

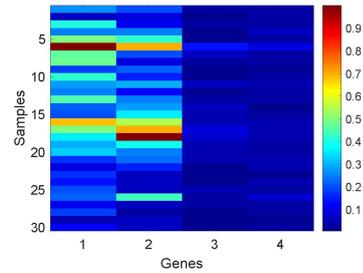

Fig. 18. Heat map of the samples for the Colon dataset. Each panel corresponds to one gene. From small to high values represents low to high expression levels of samples. The image reveals that moderate to high upregulation of H08393 and X12671 and downregulation for other two irrelative genes (M22488 and R72644).

To further study the biological function of the candidate pathogenicity-related genes, we also perform the functional enrichment analysis of the top 178 genes identified by our method on the website https://david.ncifcrf.gov/. The results of KEEG_PATHWAY are listed in Table 8. It can be seen from this table that the item of DNA replication has the lowest p-value, so it is considered as the most probable enrichment item. Some other items with the most significance are also listed in this table, for example, the first five pathways have statistical meaning ($p<0.05$). For genes enrich in these pathways, we further do Kaplan-Meier curve by anglicizing survival curves and corresponding Log-Rank $P$ values. We have found two proto-oncogenes (NCBP2 with $P=0.0217$ and ITGA7 with $P=0.0183$) and one anti-oncogene (TPM1 with $P=0.0197$). Fig. 19 shows that, for proto-oncogenes and

TABLE 8

THE KEEG PATHWAY TERMS ENRICHMENT ANALYSIS OF THE TOP 178 GENES IN THE COLON DATA SET BY DAVID

| Rank | KEEG-PATHWAY | P-value |
|---|---|---|
| 1 | DNA replication | 5.7E-3 |
| 2 | Spliceosome | 8.8E-3 |
| 3 | Hypertrophic cardiomyopathy(HCM) | 1.1E-2 |
| 4 | Dilated cardiomyopathy | 1.4E-2 |
| 5 | Arrhythmogenogenic right ventricular | 4.2E-2 |
| 06 | ECM-receptor interaction | 5.4E-2 |
| 7 | Pyruvate metabolism | 6.3E-2 |
| 8 | Aminoacy-tRNA biosynthesis | 6.5E-2 |
| 9 | Purine metabolism | 7.0E-2 |
| 10 | Pyrimidine metabolism | 7.2E-2 |

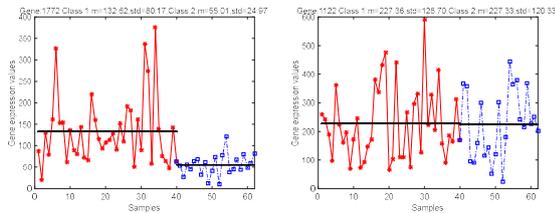

Fig. 17. Comparison of expression levels for the pathogenic genes (left) and irrelevant genes (right). For pathogenic genes H08393, the mean and standard deviation of expression levels about tumor category samples are higher than that of normal category samples But for irrelevant genes M22488, there are similar mean and std in both categories.

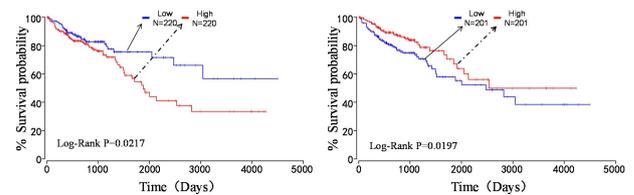

Fig. 19. Kaplan-Meier survival curves of genes which enrich in pathways with statistical meaning ($p<0.05$). Subimages from left to right: NCBP2 and TPM1 respectively. lines with dotted arrow denote upper 50% percentile and lines with solid arrow denote lower 50% percentile.

anti-oncogene, high expression and low expression have significant difference in survival rate.

## 4 CONCLUSIONS AND FUTURE WORK

In this paper, a simple, efficient and stable representation technique, IPRC, is presented for improving SRC by taking full advantages of test samples. For robust tumor classification, a two-stage hybrid gene selection algorithm is designed to combine with IPRC. Furthermore, some valuable analysis of candidate pathogenicity-related genes is given.

There remain some interesting questions. One is how to enforce some prior constraints into the IPRC model based on different applications. Another is to seek more effective gene selection methods.

## ACKNOWLEDGMENTS

The authors would like to thank http://www.gems-system.org/ for their datasets. We also thank Y. Li, P. Wang and W. Yue for bioinformatics' suggestion, Z. Zhang and Y. Shi for statistical experiments, X. Jiang, C. Tian and L. Sun for plotting. This work was supported in part by NSF of China (11701144), NSF of US (DMS1719932), NSF of Henan Province (162300410061), Key Project of the Education Department Henan Province (14A120009), and Project of Emerging Interdisciplinary of Henan University (xxjc20170003). X. Yang is the corresponding author.

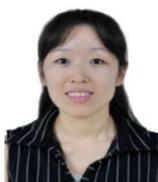

**Xiaohui Yang** received the Ph. D. degree in Institution of Intelligence Information Processing from Xi'dian University, in 2007. Since 2007, she is an associate professor with the School of Mathematics and Statistics, Henan University. She was a visiting scholar of University of Florida, USA, from 2016.8 to 2017.8, Her research interests include pattern recognition, intelligence information processing.

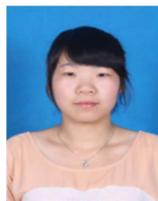

**Li Tian** received the B.S. degree in School of Mathematical Sciences from Henan Institute of Science and Technology, China, in 2014. Currently, she is studying for a master's degree at the School of Mathematics and Statistics, Henan University. Her research interests include pattern recognition, biomedical data processing.

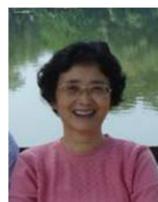

**Yunmei Chen** received the Ph.D. degree in Departments of Mathematics from Fudan University, Shanghai, China, in 1985. Since 1991, she is working in University of Florida, Gainesville, FL, USA. She is a distinguished Professor since 2015. Her current research interests include medical data analysis, optimization techniques and applications in imaging, machine learning and network computing.

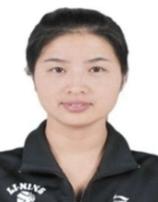

**Lijun Yang** received the Ph.D. degree in the School of Mathematics from Sun Yat-sen University, in 2013. Since 2017, she is an associate professor with the School of Mathematics and Statistics, Henan University. Her current research interests include time-frequency analysis, signal processing and pattern recognition.

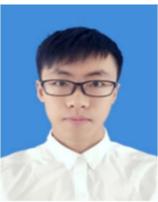

**Shuang Xu** received the B.S. degree in School of Mathematics and Statistics from Henan University, China, in 2012 and is currently a graduate student for a Master's degree at the School of Mathematics and Statistics, Xi'an Jiaotong University, Shaanxi, China. His research interests include statistics, data mining and complex network and system.

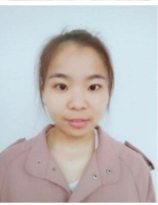

**Wenming Wu** received the B.S. degree in School of Mathematical Sciences from Henan Institute of Science and Technology, China, in 2015. Currently, she is studying for a master's degree at the School of Mathematics and Statistics, Henan University. Her research interests include pattern recognition, biomedical data processing.